# An Approach to Maintaining Safety Case Evidence After A System Change

Omar Jaradat[*], Patrick Graydon[*]
[*]School of Innovation, Design, and Engineering
Mälardalen University
Västerås, Sweden
Email: {omar.jaradat, patrick.graydon}@mdh.se

Iain Bate[†]
[†]Department of Computer Science
University of York
York, United Kingdom
Email: iain.bate@york.ac.uk

*Abstract*—Developers of some safety critical systems construct a safety case. Developers changing a system during development or after release must analyse the change's impact on the safety case. Evidence might be invalidated by changes to the system design, operation, or environmental context. Assumptions valid in one context might be invalid elsewhere. The impact of change might not be obvious. This paper proposes a method to facilitate safety case maintenance by highlighting the impact of changes.

## I. Introduction

Developers of some safety critical systems construct a safety case comprising both safety evidence (e.g. safety anal-yses, software inspections, or functional tests) and a safety argument explaining that evidence. The safety argument shows which claims the developer uses each item of evidence to support and how those claims, in turn, support broader claims about system behaviour, hazards addressed, and, ultimately, acceptable safety. Changes to the system during or after development might invalidate safety evidence or argument. Evidence might no longer support the developers' claims because it reflects old development artefacts or old assumptions about operation or the operating environment. In the updated system, existing safety claims might be nonsense, no longer reflect operational intent, or be contradicted by new data. To maintain the safety case after the system is changed, developers must analyse the change's impact. This analysis is traditionally done by hand: developers determine whether the evidence still supports the claims made of it, check to see whether new or updated safety requirements are reflected in the argument, and manually review the argument's logic. In this paper, we propose a method to facilitate safety case change impact analysis by automatically highlighting some kinds of impacts.

For the sake of simplicity, we assume in this paper that safety arguments are recorded in the Goal Structuring Notation (GSN) [2]. However, the method we propose might (with suitable adaptations) be suitable for use with other graphical assurance argument notations.

## II. Our Proposal

A complete approach to managing safety case change would include both (a) mechanisms to structure the argument so as to contain the impact of predicted changes and (b) means of assessing the impact of change on all parts of the argument. In this paper, we focus on identifying the evidence that must be updated to reflect any given change.

To facilitate identifying the evidence impacted by change, we propose storing additional information in the safety argu-ment. We propose annotating each reference to a development artefact (e.g. an architecture specification) in a goal or context element with an artefact version number. We also propose annotating each solution element with:

1) An evidence version number
2) An input manifest identifying the inputs (including version) from which the evidence was produced
3) The lifecycle phase during which the evidence ob-tained (e.g. Software Architecture Design)
4) A safety standard reference to the clause in the applicable standard (if any) requiring the evidence (and setting out safety integrity level requirements)

With this data, we can perform a number of automated checks to identify items of evidence impacted by a change. For example:

1) We can determine when two different versions of the same item of evidence are cited in the same argument
2) We can identify out-of-date evidence by searching for input manifests m = {($a_1$, $v_1$) , ..., ($a_n$, $v_n$)} and artefact versions (a, v) such that $9i • a = a_i \wedge v > v_i$
3) Where we know a particular artefact has changed, we can search for input manifests containing old versions

If we had further information which inputs were used to produce each input listed in each input manifest, each input that was used to produce those, and so on, we could extend checks (2) and (3) above to indirect inputs. For example, suppose that life testing is used to establish the reliability of a component, that this component and its reliability appear in a Failure Modes and Effects Analysis (FMEA), and that the FMEA results are used in a Fault Tree Analysis (FTA). With the additional information, we could compute a closure of the FTA's input manifest that would include the life testing results.

Other analyses may be possible. For example, we suggest storing the safety standard reference to facilitate analysis of impacts that change the safety integrity level of a requirement. However, we have not yet thought these through.

## III. An Illustrative Example

To illustrate our proposal, consider how the analysis might work on a sample system. Figure 1 presents part of an assurance argument we built for a specimen safety critical system we built in prior work [3]. The Fuel Level Estimation System (FLES) is meant to monitor fuel levels to prevent loss of engine power due to running out of fuel. (Running out of fuel is a serious problem in heavy road vehicles because steering and braking mechanisms are powered by the engine; loss of engine power while driving could result in an accident.)

The argument fragment concerns model checking anal-yses of the system architecture [3]. The FLES architec-ture, specified in the Architecture Analysis and Design Lan-guage (AADL) [1], comprises five threads: SoftwareIN, FuelEstimation, FuelLevelWarning, Other_Functions and SoftwareOUT. These threads run on a single-core mi-croprocessor with non-preemptive scheduling. Using the UP-PAAL model checker, we verified that the system as spec-ified in the architecture is schedulable and free from live-lock and deadlock. Rectangular goal element G:LivelocksFree represents the claim that the architecture is free from live-lock. G:LivelocksFree's connection to round solution element S:CtrlFloAn shows that this claim is supported by the control flow analysis done using the model checker.

The green elements in Figure 1 represent the annotations described in Section II. (These need not necessarily be pre-sented to the user in visual depictions.) Let us consider an example change scenario to illustrate how this information aids safety case change impact analysis. Suppose that the architec-ture was simplified by removing the FuelEstimation thread and moving the tasks it contains to the FuelLevelWarning thread. Suppose that an engineer making this change had up-dated the artefact version annotation(s) in part of the argument referring to the functional behaviour of in those threads. An automated implementation of check (2) described in Section II could highlight the need to re-run the control flow analysis as well. If the new version of the architecture is version 1.1, analysis of the manifest associated with S:CtrlFloAn would reveal evidence based on an older version of the architecture and tools could flag S:CtrlFloAn as out of date and suspect.

Automated analysis might also highlight goal G:Estimator-ArchFree because its artefact version annotation refers to an out-of-date version of the AADL architecture. The goal and its supporting argument are suspect because they might refer to parts of the architecture that no longer exist or make claims about the architecture that are no longer true.

## IV. RELATED WORK

Weaver, McDermid, and Kelly proposed characterising safety evidence according to, amongst other things, the type of technique that produced it (e.g., analysis, testing, inspection, etc.) [4]. Their characterisation was meant to facilitate judg-ment of the sufficiency of the evidence. We propose a different characterisation of safety evidence with a different purpose.

Tracking version information and using it to determine when artefacts are out of date is by no means new; make does this. Our contribution lies in applying this idea to safety arguments and safety case change impact analysis.

## V. CONCLUSION

Maintaing safety arguments after implementing a system change is painstaking process. In this paper we propose a new approach to facilitating safety case change impact analysis. In the approach, automated analysis of information given as annotations to the safety argument highlights suspect safety evidence to bring it to engineers' attention. We illustrated the approach with an example drawn from an automotive system.

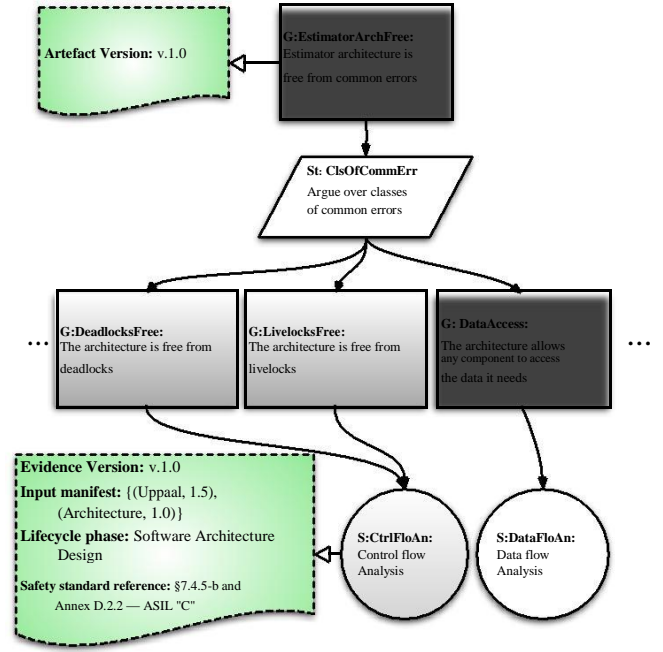

Fig. 1. Model Checking Module — Argument Fragment [3].

We have not considered the full range of properties that we could check with automated analyses or the annotations necessary to support those analyses. We have likewise not yet studied the feasibility or value of such automated checks by implementing and applying them. We leave these efforts to future work.

ACKNOWLEDGMENT

We acknowledge the Swedish Foundation for Strategic Research (SSF) SYNOPSIS Project for supporting this work.